\documentclass[aps,twocolumn]{revtex4}%
\usepackage{amsfonts}
\usepackage{amsmath}
\usepackage{amssymb}
\usepackage{epsfig}
\usepackage{graphicx}%
\ifx\pdfoutput\relax\let\pdfoutput=\undefined\fi
\newcount\msipdfoutput
\ifx\pdfoutput\undefined\else
\ifcase\pdfoutput\else
\ifx\paperwidth\undefined\else
\ifdim\paperheight=0pt\relax\else\pdfpageheight\paperheight\fi
\ifdim\paperwidth=0pt\relax\else\pdfpagewidth\paperwidth\fi
\fi\fi\fi
\begin{document}
\title{Nonsinusoidal current-phase relation in strongly ferromagnetic and moderately
disordered SFS junctions}
\author{Francois K\textsc{onschelle}}
\author{J\'{e}r\^{o}me C\textsc{ayssol} }
\author{Alexandre I. B\textsc{uzdin}}
\altaffiliation{Also at \textit{Institut Universitaire de France}}

\affiliation{Universit\'{e} de Bordeaux ; CNRS ; CPMOH, F-33405 Talence Cedex, France}
\date{{\today}}
\startpage{1}

\begin{abstract}
We study the Josephson current in a junction comprising two superconductors
linked by a strong ferromagnet in presence of impurities. We focus on a regime
where the electron (and hole) motion is ballistic over the exchange length and
diffusive on the scale of the weak link length. The current-phase relation is
obtained for both two- and three dimensional ferromagnetic weak links. In the
clean limit, the possibility of temperature-induced $0$-$\pi$ transitions is
demonstrated while the corresponding critical current versus temperature
dependences are also studied.

\end{abstract}
\maketitle

\bigskip

\bigskip

\section{Introduction.}

The Josephson effect is a striking manifestation of quantum mechanics at
macroscopic scales \cite{josephson62}. When a small current $I$ is driven
through a superconductor/insulator/superconductor junction, no voltage drop
occurs along the junction while a finite phase difference $\chi$ appears
between the two superconducting order parameters of the leads. When the
applied current exceeds a maximal (critical) value $I_{c}$, a finite voltage
appears across the barrier yielding a time-dependence of the phase. In many
cases the stationary current phase relation (CPR) is well approximated by its
first harmonic $I(\chi)=I_{1}\sin\chi$ and then the critical current is simply
given by $I_{c}=I_{1}$. Nevertheless, theory gives room to higher harmonics,
in particular at low temperatures. In fact, the only general requirement is
that $I(\chi)$ must be a $2\pi$-periodic and odd function of the phase
difference when time-reversal invariance is respected. Thus the CPR may be
expressed as a Fourier sum $I(\chi)=\sum_{m}I_{m}\sin m\chi$ where the
coefficients $I_{m\text{ }}$are related to processes whereby $m=1,2,..$ Cooper
pairs are transferred through the weak link \cite{golubov04}.

If the junction is inserted in a superconducting loop, the supercurrent is
controlled by the applied magnetic flux $\Phi$ which is directly related to
the superconducting phase difference by $\chi=2\pi\Phi/\Phi_{0}$, $\Phi
_{0}=h/2e$ being the superconducting flux quantum. In the absence of any
bias-current or magnetic flux, the equilibrium state of a tunnel Josephson
junction usually corresponds to $\chi=0$ and $I=0$. In contrast, when the
tunnel barrier contains magnetic impurities, it was predicted that the phase
difference may be equal to $\pi$ at equilibrium. This may enable a spontaneous
persistent current to flow in a loop comprising a Josephson junction in such a
$\pi$-state \cite{bulaevskii78,bulaevskii78ss}. This $\pi$-shift is related to
processes whereby electrons change their spin projection when passing through
the insulating layer \cite{kulik66,shiba69}. Unfortunately this kind of $\pi
$-state, generated by magnetic impurities in a host insulating layer, was
never observed experimentally. In contrast it was further predicted
\cite{buzdin82,buzdin82bis} and observed \cite{ryazanov01,kontos02} that a
superconductor/ferromagnetic metal/superconductor (SFS) junction also exhibits
transitions between zero and $\pi$-groundstates when the exchange energy $h$
and/or the length $L$ of the ferromagnet is varied \cite{buzdin82}. The
corresponding current-phase relation (CPR) $I(\chi)$ were analysed for pure
\cite{buzdin82} and dirty \cite{buzdin91} ferromagnets using respectively
Eilenberger and Usadel equations
\cite{kopnin,likharev79,buzdin05,lyuksyutov05}. In \ both cases, the critical
current of a SFS junction oscillates and decays when increasing the length or
the exchange energy of the ferromagnet. The oscillations of $I_{c}(L)$
originate directly from the exchange interaction which induces a finite
mismatch between the Fermi wavevectors of spin up and down electrons. Besides
these oscillations,\ scattering by magnetic and nonmagnetic disorder strongly
suppresses the critical current when $L$ is increased. In diffusive
ferromagnets, this overall decay is exponential on the typical scale $\xi
_{1}=\sqrt{D/h}$, $D$ being the diffusion constant. In contrast, in the pure
limit, a finite Josephson current may be observed up to much larger length
scales on the order of $\xi_{1}=v_{F}/T$, $v_{F\text{ }}$and $T$ being
respectively the Fermi velocity in the ferromagnet and the temperature. In
particular at zero temperature, the decay becomes a power law, namely
$I_{c}\sim L^{-1}$ in the case of a three dimensional pure ferromagnetic weak
link \cite{buzdin82}. In the absence of disorder, this critical current
suppression results from the superposition of many distinct single channel
CPRs associated with independent transverse channels. Accordingly this decay
is expected to be less severe in low dimensional ferromagnets for it
corresponds to angular averaging over quasiclassical trajectories with
different angles with respect to the junction axis.

The first evidence of the $\pi$-state, in SFS junctions, came with the
observation of the nonmonotonic behavior of the critical current as a function
of temperature \cite{ryazanov01} and of the ferromagnet length \cite{kontos02}%
. The weak ferromagnetic alloys chosen for these pioneer experiments enable to
observe the transitions in relatively large junctions within the ten
nanometers scale. Further support was provided by magnetic diffraction
patterns of DC squids comprising a SFS $\pi$-junction in one arm and a usual
tunnel junction on the other arm \cite{guichard03}. Furthermore spontaneous
persistent currents were reported in a loop interrupted by a $\pi$ junction
\cite{bauer04} and imaged in arrays of $\pi$-Josephson junctions
\cite{ryazanovNatPhys}. Finally, multiple transitions between zero and $\pi
$-groundstates were observed by varying the ferromagnetic layer thickness of a
SFS junction \cite{oboznov06,robinson07}.

Experimentally obtaining the CPR is much more difficult than simply measuring
the critical current. Only recently a few CPR experiments were implemented
successfully in the case of SFS junctions \cite{frolov04} \ and SNS junctions
\cite{dellarocca07,troeman08,strunk08}. In fact at sufficiently low
temperature, a small second harmonic of the CPR is always present both in SNS
and SFS junctions, but it is usually completely eclipsed by the large
amplitude of the first harmonic. The SFS Josephson junctions are a natural
playground to observe unambigously the second harmonic since the first one
vanishes at the zero-$\pi$ transition. In particular, the second harmonic was
detected as a tiny minimum supercurrent at the crossover between the zero and
$\pi$ states, and also revealed by the related Shapiro steps \cite{sellier04}.
In the highly transparent limit, the issue of the sign and magnitude of the
second harmonic was adressed in presence of uniaxial \cite{buzdin05prb} and
isotropic \cite{houzet05} magnetic scattering in the ferromagnet. Moreover
finite transparency or weak interfacial disorder may also modify the
second-harmonic \cite{zareyan06}.

Zero-$\pi$ transitions were also observed in smaller junctions comprising
strong ferromagnets like Fe, Co, Ni or permalloy
\cite{blum02,robinson06,robinson07}. These novel experiments are performed in
an interesting regime which differs both from the pure clean or dirty limits
extensively studied so far. Owing to the extremelly large exchange energy, the
period of the critical current oscillations $\xi_{2}=v_{F}/h$ is smaller than
the mean free path $\ell=v_{F}\tau$ while the ferromagnetic bridge is still
longer than the mean free path: $\xi_{2}\ll\ell\ll L$. Bergeret \textit{et
al.} investigated the first harmonic of the Josephson current in this
particular regime $h\tau\gg1$ \cite{bergeret01}. A theoretical analysis of
this second harmonic in this particular regime is still lacking while it was
already detected experimentally \cite{robinson07}.

\begin{figure}[ptb]
\begin{center}
\includegraphics[height=3in,angle=270]{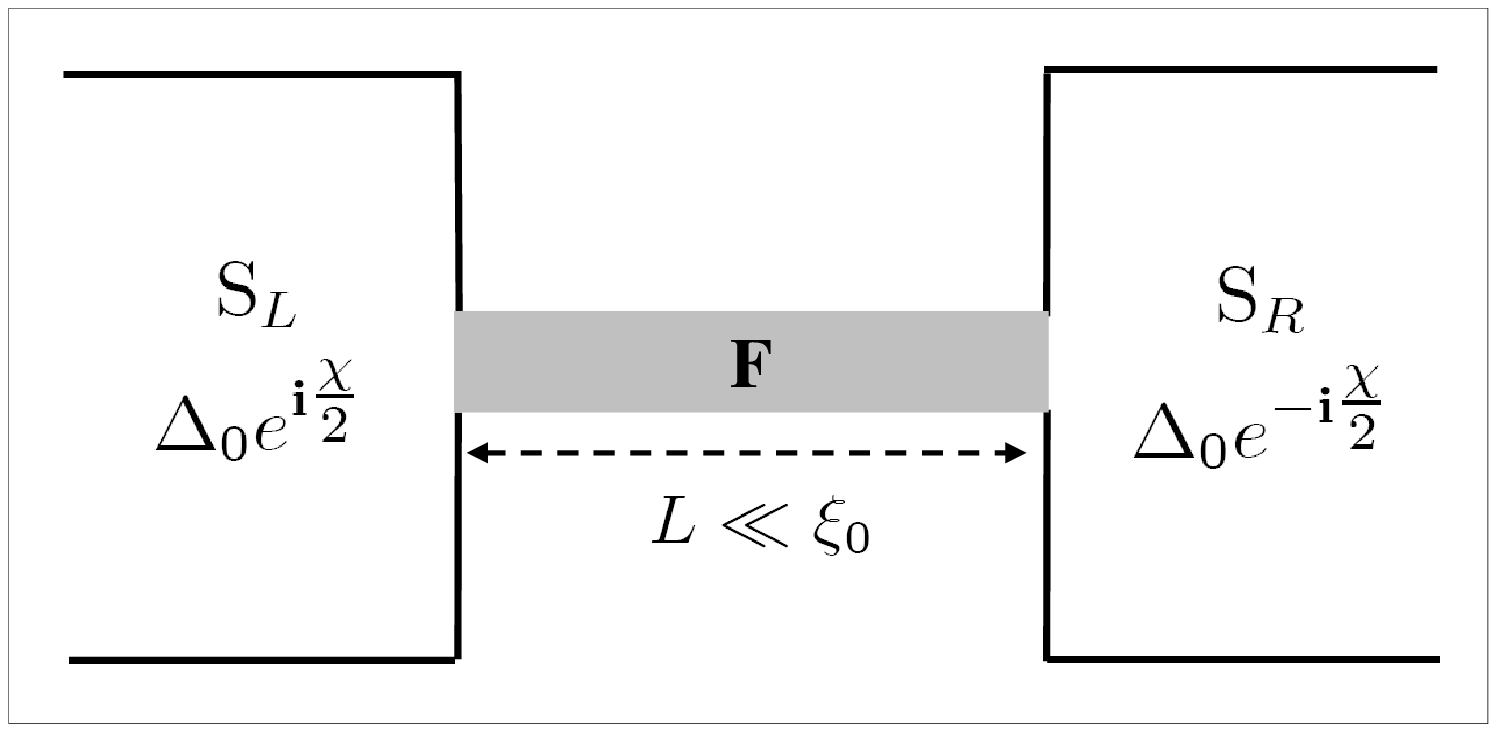}
\end{center}
\caption{SFS junction. The contact area are far smaller than the typical
superconducting lead ($S_{L},S_{R}$) transverse size. The metallic ferromagnet
(F) in the middle part can be either two- or three- dimensional.}%
\label{FIG_sketch_junction}%
\end{figure}

Up to now the physics of the $\pi$-state and second harmonic were mostly
investigated in the three-dimensional case due to the lack of
lower-dimensional ferromagnets. Recenlty novel systems like graphene or thin
films of magnetic semiconductors became available as promising candidates for
realizing two-dimensional SFS junctions. For instance coating graphene with Pd
may produce itinerant magnetism \cite{uchoa08} while alcaline coating is
likely to produce superconductivity \cite{uchoa07}. Hence tailoring a graphene
sheet with appropriate metals on top may induce SFS heterojunctions within the
carbon atoms plane. At the present time, only two experiments have been
reported on SNS\ junctions made with graphene \cite{heersche07,du08}, while
$0$-$\pi$ transitions were predicted in graphene based SFS junctions
\cite{huertas08}. Another experiment which has triggered the interest on
two-dimensional SFS junctions is the measurement of a supercurrent through a
long, $L=0.3-1\mu m$, half-metallic ferromagnet chromium oxide (CrO$_{2\text{
}}$) film \cite{keiser06} wherein singlet superconductivity should be
destroyed on a much smaller length scale. The spin singlet to triplet
conversion (at the interfaces) was proposed to explain the finite Josephson
current. It was shown that triplet correlations penetrate a ferromagnet on
much longer distances \cite{bergeret05}. Another possibility is that the
surface of the film is less spin-polarized than the bulk or even
antiferromagnetically ordered. In this latter scenario, a singlet supercurrent
may bypass the half-metallic ferromagnet by flowing within a two-dimensional
SFS surface junction.

In this paper, we study both two- and three-dimensional SFS junctions with
strong exchange field and moderate disorder, namely in the limit $h\tau\gg1$.
Using Eilenberger equations and perturbative expansion in $1/h\tau$, we obtain
the current phase relation and in particular its second-harmonic at the
$0$-$\pi$ transition which was recently observed in the three-dimensional case
\cite{robinson07}. In the pure limit, we show that the two-dimensional
critical current is suppressed as $L^{-1/2}$ instead of $L^{-1}$ in the
three-dimensional case. Accordingly we suggest the possibility to observe
enhanced critical current in planar SFS junctions made of magnetic
semiconductors or graphene with an induced ferromagnetic order.

After introducing the formalism in Sec. II, we investigate the pure limit in
Sec. III with special emphasis on the two-dimensional case and on
temperature-induced $0$-$\pi$ transitions. In Sec IV, the CPR of two- and
three-dimensional SFS junctions are obtained in the limit $h\tau\gg1$ in
relation with experiments \cite{robinson07}.

\bigskip

\section{SFS model and formalism.\label{PAR_model_formalism}}

We study a superconductor-ferromagnetic-superconductor (SFS) Josephson
junction in the geometry represented in Fig.\ref{FIG_sketch_junction}. The
ferromagnet consists in a single ferromagnetic domain characterized by its
exchange energy $h$, length $L=L_{x}$ and transverse dimension(s) $L_{y}$ (and
$L_{z}$). We assume that the superconducting order parameter is $\Delta
_{0}e^{i\chi/2}$ (resp. $\Delta_{0}e^{-i\chi/2}$) in the left (right)
superconducting lead while the Fermi velocity $v_{F}$ is the same everywhere.
The contacts between superconductors and the ferromagnet are completely
transparent and spin inactive. Besides the geometrical parameters of the
junction, three typical lengths are of primary importance for the Josephson
effect. On the one hand the superconducting coherence length $\xi_{0}%
=v_{F}/\pi\Delta_{0}$ and the ferromagnetic exchange length $\xi_{F}=v_{F}/h$
are related to the strengh of the superconducting and ferromagnetic order
parameters respectively. On the other hand disorder is characterized by the
elastic mean free path $\ell=v_{F}\tau$ where $\tau$ is the average time
between impurity scattering events. Henceforth, we adopt units with
$\hbar=k_{B}=1$.

Previous theoretical studies were mostly performed in the diffusive limit
$\ell\ll\xi_{F},L$ or in the pure limit $\ell\gg\xi_{F},L$ using respectively
the Usadel and the Eilenberger equations (without self-energy terms due to
disorder) \cite{kopnin}. Recent experiments have opened to possibility to
investigate the regime $\xi_{F}\ll\ell\ll L$
\cite{blum02,robinson06,robinson07}. This later regime cannot be described by
the Usadel equation since the disorder induced self-energy is no longer the
dominant energy. It is thus necessary to use the Eilenberger formalism
\cite{bergeret01} \ and the exchange energy as the large parameter which
enable perturbative expansions.

In our simple model, the quasi-classical Eilenberger Green functions
$g=g_{\omega}\left(  x,v_{x}\right)  $, $f=f_{\omega}\left(  x,v_{x}\right)  $
and $f^{+}=f_{\omega}^{+}\left(  x,v_{x}\right)  $ depend only on the
center-of-mass coordinate $x$ along the junction axis $Ox,$ and on the angle
$\theta$ of the quasiclassical trajectories with respect to $Ox$.

In the ferromagnetic weak link, $\left\vert x\right\vert <L/2$, the
Eilenberger equations read
\begin{equation}
\left\{
\begin{array}
[c]{l}%
v_{x}\partial_{x}g=\left(  2\tau\right)  ^{-1}\left(  f\left\langle
f^{+}\right\rangle -f^{+}\left\langle f\right\rangle \right)  ,\\
v_{x}\partial_{x}f=-2\left(  \omega+ih\right)  +\dfrac{1}{\tau}\left(
g\left\langle f\right\rangle -f\left\langle g\right\rangle \right)  ,\\
-v_{x}\partial_{x}f^{+}=-2\left(  \omega+ih\right)  f^{+}+\dfrac{1}{\tau
}\left(  g\left\langle f^{+}\right\rangle -f^{+}\left\langle g\right\rangle
\right)  .
\end{array}
\right.  \label{EQ_eilenberger_ferro}%
\end{equation}
Here $\omega=\pi T\left(  2n+1\right)  $ are the Matsubara frequencies and
$v_{x}=v_{F}\cos\theta$ is the Fermi velocity vector \cite{kopnin}. The
brackets denote averaging over the Fermi surface.

In the superconducting leads, which are assumed to be clean, the Eilenberger
equations read :
\begin{equation}
\left\{
\begin{array}
[c]{l}%
v_{x}\partial_{x}g=\Delta^{\ast}f-\Delta f^{+}\\
v_{x}\partial_{x}f=-2\omega f+2\Delta g\\
v_{x}\partial_{x}f^{+}=2\omega f^{+}-2\Delta^{\ast}g
\end{array}
\right.  \label{EQ_eilenberger_supra}%
\end{equation}
with $\Delta=\Delta_{0}e^{i\chi/2}$ (resp. $\Delta_{0}e^{-i\chi/2}$) for the
left (resp. right) electrode respectively. In the whole paper, $\Delta
_{0}=1,764T_{c}\tanh\left(  1,74\sqrt{T/T_{c}-1}\right)  $ is the temperature
dependent superconducting gap.

In the limit $h\tau\gg1$ studied thorough this paper, one may assumes as a
starting approximation that $\left\langle f\right\rangle =\left\langle
f^{+}\right\rangle =0$. Then demanding the continuity of the general solutions
of Eqs. ($\ref{EQ_eilenberger_ferro},\ref{EQ_eilenberger_supra})$ at the
interfaces $x=\pm L/2$ yields the quasiclassical Green functions over the
whole junction. In particular, in the ferromagnet, $\left\vert x\right\vert
<L/2$, one finds that the normal Green function
\begin{equation}
g\left(  x,v_{x}\right)  =\frac{\omega}{\Omega}+\frac{\Delta_{0}^{2}}{\Omega
}\frac{\sinh\Phi}{\omega\sinh\Phi\pm\Omega\cosh\Phi} \label{EQ_g}%
\end{equation}
is independent of the position $x$. The upper (lower) sign of the denominator
corresponds to the positive (negative) sign in the velocity projection. We
have defined $\Omega^{2}=\omega^{2}+\Delta_{0}^{2}$ and the effective phase
\begin{equation}
\Phi=\dfrac{\omega L}{v_{x}}+\frac{\left\langle g\right\rangle L}{2v_{x}\tau
}+i\left(  \dfrac{hL}{v_{x}}+\dfrac{\chi}{2}\right)
\label{EQ_effective_phase}%
\end{equation}
which contains all the relevant parameters of the junction. In experiments,
the exchange field $h$ is always larger than the temperature, and thus the
contribution $\omega L/v_{x}$ may be safely neglected in\ the above expression
for $\Phi$.

\bigskip The supercurrent density is given by the following quasiclassical
expression \cite{kopnin}
\begin{equation}
j\left(  \chi\right)  =2\pi e\nu_{0}^{\left(  d\right)  }T\sum_{\omega
}\left\langle v_{x}\operatorname{Im}g_{\omega}\left(  v_{x}\right)
\right\rangle \label{EQ_current_definition}%
\end{equation}
where the temperature $T$ is expressed in energy units and $\nu_{0}^{\left(
d\right)  }$ is the density of state at the Fermi level per spin and per unit
volume (resp. surface) for $d=3$ (resp. $d=2$). The corresponding current
$\ I\left(  \chi\right)  $ is obtained as the flux of $j\left(  \chi\right)  $
through a section of the weak link.

Finally the groundstate energy $E_{\chi}$ of the junction can be deduced by
integrating the CPR according to the general formula%
\begin{equation}
I\left(  \chi\right)  =\frac{2\pi}{\Phi_{0}}\frac{\partial E_{\chi}}%
{\partial\chi}, \label{EQ_def_energy}%
\end{equation}
where $\Phi_{0}=h/2e$ is the superconducting flux quantum. The $0-\pi$ phase
transition occurs when the $\chi=0$ and $\chi=\pi$ groundstates are
degenerate, namely when $E_{0}=E_{\pi}$.

\section{SFS junction in the pure limit.\label{PAR_without_tau}}

In this section, we consider the pure limit $L\ll\ell$ with special emphasis
on the two-dimensional case. Indeed, the three-dimensional and the
one-dimensional cases are well known for both small \cite{buzdin82} and large
\cite{cayssol04,cayssol05} exchange energies. After briefly recalling the
single channel results, we obtain that the low temperature critical current of
a two-dimensional Josephson junction decays as $L^{-1/2}$, namely more slowly
than the $L^{-1}$ dependence characterizing three-dimensional ballistic weak
links. We obtain the CPR and study the second harmonic at the $0$-$\pi$
transition, where the first harmonic cancels. We also study the possibility of
$\ 0$-$\pi$ transitions induced by varying the temperature at a given length.
Finally, the $I_{c}(T)$ curves are obtained and compared to recent experiments
\cite{oboznov06,robinson07}.

\subsection{Single channel case.}

State of the art ferromagnetic wires typically still contain a large number of
transverse channels. Nevertheless, the analysis of the single channel case is
both necessary and instructive since, in the ballistic limit, it is the
building block for evaluating the multichannel supercurrent in higher
dimensions. When considering a single transverse channel, the angular
averaging $\left\langle ..\right\rangle $ in Eq.$\left(
\ref{EQ_current_definition}\right)  $ reduces to a discrete sum over
$\theta=0$ and $\theta=\pi$ which yields the following current-phase
relation:
\begin{equation}
I\left(  \chi\right)  =I_{0}^{\left(  1\right)  }\sum_{\sigma=\pm1}%
\tanh\left(  \dfrac{\Delta_{0}}{2T}\cos\dfrac{\chi+\sigma\alpha}{2}\right)
\sin\dfrac{\chi+\sigma\alpha}{2}, \label{EQ_monocanal_current}%
\end{equation}
where $\alpha=2hL/v_{F}$ and $I_{0}^{\left(  1\right)  }=e\nu_{0}^{\left(
1\right)  }\pi v_{F}\Delta_{0}$. Using Eq.$\left(  \ref{EQ_def_energy}\right)
$, one obtains the energy of the junction
\begin{equation}
\frac{E_{\chi}\left(  \alpha,T\right)  }{E_{0}^{\left(  1\right)  }}%
=-\sum_{\sigma=\pm1}\ln\left[  \cosh\left(  \dfrac{\Delta_{0}}{2T}\cos
\dfrac{\chi+\sigma\alpha}{2}\right)  \right]  \label{EQ_energy_1D}%
\end{equation}
as a function of the phase difference $\chi$. The typical energy scale is
given by $E_{0}^{\left(  1\right)  }=2T\pi\nu_{0}^{\left(  1\right)  }\hslash
v_{F}$. The $\chi=0$ and $\pi$ groundstates are degenerate for regularly
spaced values of the accumulated phase $\alpha$, namely at $\alpha_{n}%
=\pi/2+n\pi$. These critical values of the parameter $\alpha$ can be reached
by varying either the length $L$ or the exchange energy $h,$ whereas they are
insensitive to temperature variation. Hence in an hypothetical single channel
SFS junction, it would be impossible to drive the $0$-$\pi$ transition by
varying the temperature only.

Close to the critical temperature, $T\approx T_{c}$, the linearization of
Eq.$\left(  \ref{EQ_monocanal_current}\right)  $ yields a nearly sinusoidal
current-phase relation $I\left(  \chi\right)  =I_{c}(\alpha)\sin\chi$ whose
critical current, $I_{c}(\alpha)=I_{0}^{\left(  1\right)  }(\Delta_{0}%
/2T_{c})\cos\alpha$, cancels for $\alpha=\pi/2+n\pi$, namely at the $0$-$\pi$
transitions. Moreover the oscillatory behavior of $I_{c}(\alpha)$ is not
damped when $\alpha,$ or equivalently $L$ and/or $h$, are increased. This
behaviour is due to the absence of angular averaging in the single channel situation.

At lower temperatures, $T\ll T_{c}$, the CPR becomes nonsinusoidal. The
corresponding critical current $I_{c}\left(  \alpha\right)  $ is obtained
numerically by maximazing the current density $I\left(  \chi\right)  $ given
by Eq.$\left(  \ref{EQ_monocanal_current}\right)  $. This critical current
$I_{c}(\alpha)$ also exhibits periodic oscillations as a function of
$\alpha=2hL/v_{F}$. In contrast to the situation for $T\approx T_{c}$, the
current $I_{c}\left(  \alpha\right)  $ is finite at the cusps owing to the
presence of a sizeable second harmonic. Finally, it is very instructive to
check the sign of this second harmonic $I_{2}(\alpha_{n})$ at the zero-$\pi$
transitions where first harmonic cancels $I_{1}(\alpha_{n})=0,$ since this
sign is related to the order of the transition. It turns out that
$I_{2}(\alpha_{n})>0$ which yields a discontinuous (first-order) phase
transition between the $0$ to the $\pi$ phases. Otherwise, namely for
$I_{2}(\alpha_{n})<0$, the transition would have been continuous
(second-order) with a groundstate corresponding to an arbitrary value of the
phase difference, distinct from $0$ or $\pi$ \cite{buzdin05}.

\subsection{Two-dimensional case.}

\begin{figure}[ptb]
\begin{center}
\includegraphics[width=3.5in]{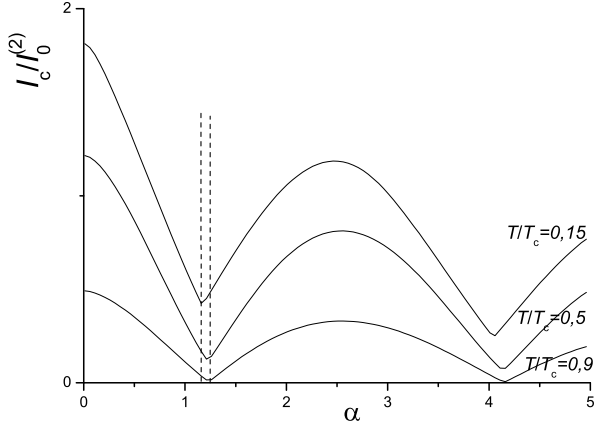}
\end{center}
\caption{Critical current as a function of $\alpha=2hL/v_{F}$ in a
two-dimensional SFS\ junction. The temperature is increased from the upper
curve ($T\approx0$) to the lower one ($T\approx T_{c}$). \ The $0-\pi$ phase
transitions occur at the kinks of the $I_{c}(\alpha)$ curve. For a given
$0-\pi$ transition, the positions of $\alpha_{n}(T)$ of these kinks exhibit
weak temperature dependence.}%
\label{FIG_critical_current_vs_thickness_2D}%
\end{figure}

In a two-dimensional SFS junction, the supercurrent is the sum of the currents
carried by independent transverse modes. Using the angular averaging
$\left\langle ..\right\rangle =\int d\theta/2\pi(..)$ appropriate for planar
junctions$,$ one obtains the following CPR:
\begin{multline}
I\left(  \chi\right)  =I_{0}^{\left(  2\right)  }\sum_{\sigma=\pm1}%
\int_{\alpha}^{\infty}\frac{\alpha^{2}dy}{y^{2}\sqrt{y^{2}-\alpha^{2}}}%
\times\\
\times\tanh\left(  \frac{\Delta_{0}}{2T}\cos\frac{\chi+\sigma y}{2}\right)
\sin\frac{\chi+\sigma y}{2} \label{EQ_current_phase_2D}%
\end{multline}
where $\alpha=2hL/v_{F}$ and $I_{0}^{\left(  2\right)  }=e\nu_{0}^{\left(
2\right)  }\pi v_{F}\Delta_{0}$. The corresponding critical current
$I_{c}\left(  \alpha\right)  $ is shown in
Fig.\ref{FIG_critical_current_vs_thickness_2D} for several temperatures. In
contrast to the single channel case, the oscillations of $I_{c}(\alpha)$ are
damped due to the angular averaging over many transverse channels having each
a distinct CPR. For $T<T_{c}$,\ the curves $I_{c}(\alpha)$ exhibit cusps where
the critical current remains finite instead of the cancellations observed for
$T\approx T_{c}$.

The zero- and $\pi$-groundstate energies $E_{0}(\alpha)$ and $E_{\pi}(\alpha)$
depend both of $\alpha$ and temperature according to:
\begin{align}
\frac{E_{0}\left(  \alpha\right)  }{E_{0}^{\left(  3\right)  }}  &
=-2\int_{\alpha}^{\infty}\ln\left[  \cosh\left(  \frac{\Delta_{0}}{2T}%
\cos\frac{y}{2}\right)  \right]  \frac{\alpha^{2}dy}{y^{2}\sqrt{y^{2}%
-\alpha^{2}}},\\
\frac{E_{\pi}\left(  \alpha\right)  }{E_{0}^{\left(  3\right)  }}  &
=-2\int_{\alpha}^{\infty}\ln\left[  \cosh\left(  \frac{\Delta_{0}}{2T}%
\sin\frac{y}{2}\right)  \right]  \frac{\alpha^{2}dy}{y^{2}\sqrt{y^{2}%
-\alpha^{2}}}.
\end{align}

The values of the parameter $\alpha=2hL/v_{F}$ where the $\left(  0\text{-}%
\pi\right)  $ transitions occur are obtained by solving $E_{0}(\alpha)=E_{\pi
}(\alpha)$ which may be rewritten as
\begin{equation}
\int_{\alpha_{c}}^{\infty}\ln\frac{\cosh\left(  \dfrac{\Delta_{0}}{2T}%
\cos\dfrac{y}{2}\right)  }{\cosh\left(  \dfrac{\Delta_{0}}{2T}\sin\dfrac{y}%
{2}\right)  }\frac{dy}{y^{2}\sqrt{y^{2}-\alpha_{c}^{2}}}=0. \label{alpha2d}%
\end{equation}

At low temperature $T<T_{c},$ we have checked that the solutions $\alpha
_{c}(T)$ for Eq.(\ref{alpha2d}) coincide with the location of the cusps in the
$I_{c}\left(  \alpha\right)  $ curves. Moreover, $\alpha_{c}(T)$ decreases
when the temperature is lowered, see Fig.
\ref{FIG_critical_current_vs_thickness_2D}. The phase diagram in the $\alpha
$-$T$ plane is similar than the three dimensional phase diagram shown in
Fig.\ref{FIG_phase_diagram}. It should be also emphasized that this transition
is not accompanied by a non-monotonic behavior of the $I_{c}\left(  T\right)
$ curves in contrast to the dirty case \cite{oboznov06}. \ 

\begin{figure}[ptb]
\begin{center}
\includegraphics[width=3.5in]{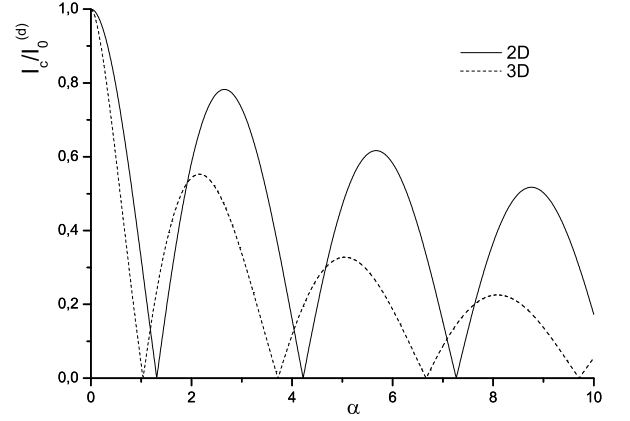}
\end{center}
\caption{Critical current $I_{c}\left(  \alpha\right)  $ versus $\alpha
=2hL/v_{F}$ for $T\approx T_{c}$ , from Eq.$\left(
\ref{EQ_critical_current_near_Tc_2D}\right)  $ in the two-dimensional case
(solid curve) and from Eq.$\left(  \ref{EQ_critical_current_near_Tc_3D}%
\right)  $ in the three-dimensional case (dashed curve). In the
two-dimensional case, the period of $I_{c}$ is larger than in the
three-dimensional case. The overall decay of $I_{c}$ is also much slower in
the two dimensional case. }%
\label{FIG_Ic_versus_alpha_near_Tc_2D_3D}%
\end{figure}

Near the critical temperature $T_{c},$ the current-phase relation is nearly
sinusoidal with a first harmonic given by
\begin{equation}
I_{1}(\alpha)=I_{0}^{\left(  2\right)  }\dfrac{\Delta_{0}}{2T_{c}}\int
_{\alpha}^{\infty}dy\frac{\alpha^{2}\cos y}{y^{2}\sqrt{y^{2}-\alpha^{2}}},
\label{EQ_critical_current_near_Tc_2D}%
\end{equation}
and a second harmonic given by:
\begin{equation}
I_{2}(\alpha)\approx-\frac{I_{0}^{\left(  2\right)  }}{12}\left(
\dfrac{\Delta_{0}}{2T}\right)  ^{3}\int_{\alpha}^{\infty}dy\frac{\alpha
^{2}\cos2y}{y^{2}\sqrt{y^{2}-\alpha^{2}}}.
\end{equation}
Both $I_{1}(\alpha)$ and $I_{1}(\alpha)$ exhibit an oscillatory dependence on
$\alpha$, see Fig.\ref{FIG_Ic_versus_alpha_near_Tc_2D_3D}. In the limit
$\alpha\gg1$, the asympotic behaviors are given respectively by
\begin{equation}
\frac{I_{1}(\alpha,T\rightarrow T_{c})}{I_{0}^{\left(  2\right)  }}%
\approx\dfrac{\Delta_{0}}{2T_{c}}\frac{\sqrt{\pi}}{2}\frac{\cos\alpha
-\sin\alpha}{\sqrt{\alpha}} \label{EQ_alpha_grand_2D_clean}%
\end{equation}
and by:
\begin{equation}
\frac{I_{2}(\alpha,T\rightarrow T_{c})}{I_{0}^{\left(  2\right)  }}%
\approx-\left(  \dfrac{\Delta_{0}}{2T_{c}}\right)  ^{3}\frac{\sqrt{\pi}}%
{24}\frac{\cos2\alpha-\sin2\alpha}{\sqrt{2\alpha}}%
\end{equation}
Thus in the regime $\alpha\gg1$, the $0$-$\pi$ transitions occur for the
values $\alpha=\pi/4+n\pi$ where the first harmonic of the CPR cancels.
Moreover the second harmonic is positive at these transitions.

\subsection{Three-dimensional case.}

\begin{figure}[ptb]
\begin{center}
\includegraphics[width=3.5in]{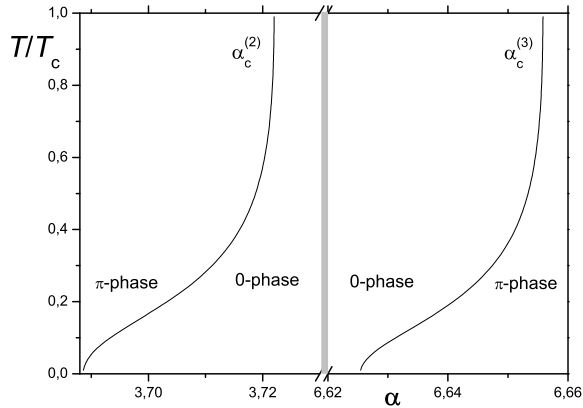}
\end{center}
\caption{Phase diagram in the $\alpha$-$T$ plane showing the transition lines
between the $0$ and the $\pi$ phases for a three dimensional SFS junction. We
have used $\Delta_{0}\left(  T\right)  =1,764T_{c}\tanh\left(  1,74\sqrt
{T/T_{c}-1}\right)  $. }%
\label{FIG_phase_diagram}%
\end{figure}

Finally we consider the well-known CPR for three-dimensional SFS junctions
\cite{buzdin82}:
\begin{multline}
I\left(  \chi\right)  =I_{0}^{\left(  3\right)  }\sum_{\sigma=\pm1}%
\int_{\alpha}^{\infty}\frac{\alpha^{2}dy}{y^{3}}\times\\
\times\tanh\left(  \frac{\Delta_{0}}{2T}\cos\frac{\chi+\sigma y}{2}\right)
\sin\frac{\chi+\sigma y}{2} \label{EQ_current_without_impurities}%
\end{multline}
in order to compare with the two-dimensional results described in the previous
paragraph. Here $I_{0}^{\left(  3\right)  }=e\nu_{0}^{\left(  3\right)  }%
L_{y}L_{z}\pi v_{F}\Delta_{0}$ and $E_{0}^{\left(  3\right)  }=2T\pi\nu
_{0}^{\left(  3\right)  }\hslash v_{F}$. The zero- and $\pi-$groundstate
energies $E_{0}(\alpha)$ and $E_{\pi}(\alpha)$ depend both of $\alpha$ and
temperature according to:
\begin{align}
\frac{E_{0}\left(  \alpha\right)  }{E_{0}^{\left(  3\right)  }}  &
=-2\int_{\alpha}^{\infty}\ln\left[  \cosh\left(  \frac{\Delta_{0}}{2T}%
\cos\frac{y}{2}\right)  \right]  \frac{\alpha^{2}dy}{y^{3}},\\
\frac{E_{\pi}\left(  \alpha\right)  }{E_{0}^{\left(  3\right)  }}  &
=-2\int_{\alpha}^{\infty}\ln\left[  \cosh\left(  \frac{\Delta_{0}}{2T}%
\sin\frac{y}{2}\right)  \right]  \frac{\alpha^{2}dy}{y^{3}}.
\end{align}
By solving numerically $E_{0}\left(  \alpha\right)  =E_{0}\left(
\alpha\right)  $ at arbitrary temperature, we obtain the curves $\alpha
_{c}(T)$ shown in Fig.\ref{FIG_phase_diagram}. In principle, the system can
experience $0-\pi$ phase transition by lowering the temperature at some given
$\alpha$, provided this value of $\alpha$ is close to a critical value. It is
important to note that the range wherein the $\left(  0-\pi\right)  $ phase
transition can take place by tuning the temperature is smaller for the
three-dimensional case than in the two-dimensional one.\ 

\begin{figure}[ptbptb]
\begin{center}
\includegraphics[width=3.5in]{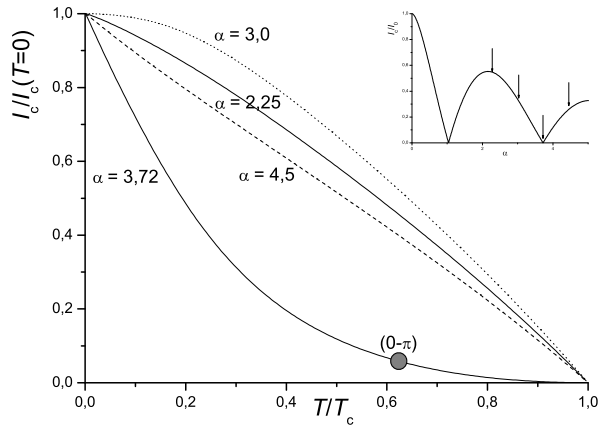}
\end{center}
\caption{Critical current $I_{c}\left(  T\right)  $ versus temperature for a
three-dimensional junction. Four different values of $\alpha=2hL/v_{F}$ are
shown: $\alpha=2.25$ (solid line), $\alpha=3.0$ (dotted line), $\alpha=3.72$
(solid line) and $\alpha=4,5$ (dashed line). The critical current decreases
monotonically when the temperature is increased. The dot on the $\alpha=3.72$
curve indicates the $\left(  0\text{-}\pi\right)  $ transition, which is not
associated with any singular behavior in the $I_{c}\left(  T\right)  $ curve.
In the inset, the above values of $\alpha$ are indicated on the $I_{c}\left(
\alpha,T\rightarrow T_{c}\right)  $ curve.}%
\label{FIG_Ic_vs_T_3D}%
\end{figure}Near the critical temperature $T_{c},$ the first harmonic is given
by%
\begin{equation}
I_{c}(\alpha)=I_{0}^{\left(  3\right)  }\dfrac{\Delta_{0}}{2T_{c}}\int
_{\alpha}^{\infty}\cos y\frac{\alpha^{2}dy}{y^{3}}
\label{EQ_critical_current_near_Tc_3D}%
\end{equation}
which is plotted on Fig.\ref{FIG_Ic_versus_alpha_near_Tc_2D_3D}. The asympotic
behaviors are given by
\begin{equation}
I_{1}(\alpha,T\rightarrow T_{c})\approx-I_{0}^{\left(  3\right)  }%
\dfrac{\Delta_{0}}{2T_{c}}\frac{\sin\alpha}{\alpha}
\label{EQ_alpha_grand_3D_clean}%
\end{equation}
in the limit $\alpha\gg1$.The second harmonic%
\begin{equation}
I_{2}(\alpha,T\rightarrow T_{c})\approx\frac{I_{0}^{\left(  3\right)  }}%
{12}\left(  \dfrac{\Delta_{0}}{2T_{c}}\right)  ^{3}\frac{\sin2\alpha}{2\alpha
}\text{ \ \ }%
\end{equation}
also exhibits an oscillatory dependence with respect to $\alpha$. Thus in the
regime $\alpha\gg1$ and close to $T_{c}$, the $0$-$\pi$ transitions occur for
the values $\alpha_{n}=n\pi$ where the first harmonic of the CPR cancels. The
second harmonic is positive at those points, $I_{2}(n\pi)>0$, indicating first
order transitions.

\subsection{Temperature dependence $I_{c}(T)$}

For appropriate values of the ferromagnetic layer length and exchange energy
(corresponding to $\alpha=\alpha_{n}$), it is possible to pass through the
$0-\pi$ \ transition point by changing the temperature, as shown in Figs.4 and
\ref{FIG_Ic_vs_T_3D} for $\alpha=3.72.$ This kind of temperature induced
$0-\pi$ transition was actually achieved in experiments using dirty weakly
ferromagnetic alloys \cite{oboznov06}. It was observed that the critical
current exhibits a nonmonotonic $T$-dependence with a cancellation at the
$0-\pi$ transition. Moreover the $\pi$ state can be either the low (e.g. at
the first node) or the high temperature phase (at second node)
\cite{oboznov06}. In contrast, only monotonic variations of $I_{c}(T)$ were
reported in experiments with strong ferromagnets in the clean (or moderately
dirty) limit \cite{robinson07,blum02}.

Here we have obtained that, in the pure limit, the critical current does
decrease monotonously when the temperature is increased, as shown in
Fig.\ref{FIG_Ic_vs_T_3D}. Nevertheless the temperature dependence of the
critical current $I_{c}(T)$ exhibits very distinct shapes (e.g. in
Fig.\ref{FIG_Ic_vs_T_3D} for $\alpha=3$ and $\alpha=3.72$) being either
convexe, concave or almost linear depending on the value of $\alpha=2Lh/v_{F}%
$. Although we have shown extreme cases (in Fig.\ref{FIG_Ic_vs_T_3D} for
$\alpha=3$ and $\alpha=3.72$), most of our $I_{c}(T)$ curves are almost linear
in agreement with the experimental curves reported in \cite{robinson07,blum02}%
. For future experiments, we suggest the observation of the concave curves
(e.g. $\alpha=3.72$ in Fig.\ref{FIG_Ic_vs_T_3D}) as a signature of the
$0$-$\pi$ transition. The corresponding measure is quite challenging since it
corresponds to the temperature dependence of a minimum of the critical current
which is given only by the contributions of higher harmonics ($m\geqslant2$).
Similar features were predicted in the limit of large exchange fields using
Bogoliubov-de Gennes formalism \cite{cayssol05}. Here we confirm that this
change in the concavity of the\ $I_{c}(T)$ curve near a $0$-$\pi$ transition
is still predicted in the limit of moderate exchange fields in comparison to
the Fermi energy. In contrast, the $I_{c}(T)$ curves are non monotonic in the
dirty limit when the junction passes the $0$-$\pi$ transition. We explain this
discrepancy by the fact that the critical current at the $0$-$\pi$ transition
vanishes in the dirty limit whereas it is still finite in the pure limit (due
to the important contribution of high harmonics).

\section{SFS junction with impurities.\label{PAR_with_tau}}

Experiments on SFS junctions comprising dilute magnetic alloys are correctly
described within the Usadel equation framework
\cite{oboznov06,buzdin05prb,houzet05,zareyan06,cottet05}, because the exchange
energy is smaller than the disorder level broadening, namely $\tau h\ll1$, and
far smaller than the Fermi energy. In this regime the electron (and hole)
motion is diffusive with a mean free path smaller than both $\xi_{F}$ and $L$.
Recently experiments were performed in the opposite regime, $\tau h\gtrsim1$,
using strong ferromagnets, like Fe, Co, Ni or Permalloy. Then the electron
(and hole) motion is ballistic over the ferromagnetic length scale $\xi_{F}$,
while being still diffusive on the scale of the weak link length $L$. In
particular this situation implies that the parameter $\alpha=2L/\xi_{F}$ is
very large. The first harmonic of the CPR for three dimensional weak links has
been already found \cite{buzdin82bis,bergeret01} by solving the Eilenberger
equations for large $\tau h$. In this section, we calculate the amplitudes of
the first and second harmonic both in two- and three- dimensional SFS
junctions. The analytical expressions obtained here can be used as a starting
point to interpret the three-dimensional experiments by Robinson \textit{et
al.} \cite{robinson07} and future investigations on graphene-based
SFS\ junctions \cite{huertas08}.

\bigskip

\subsection{Three dimensional case.}

Here we investigate the CPR $\ I(\chi)=I_{1}\sin\chi+I_{2}\sin2\chi+...$ of a
three dimensional junction. We have obtained the first harmonic
\begin{equation}
I_{1}=8\text{ }I_{0}^{\left(  3\right)  }T\sum_{\omega>0}\frac{\Delta_{0}%
}{(\omega+\Omega)^{2}}\operatorname{Re}E_{3}(z), \label{I1dirtyexact}%
\end{equation}
and the second harmonic
\begin{equation}
I_{2}=8I_{0}^{\left(  3\right)  }T\sum_{\omega>0}\frac{\Delta_{0}^{3}}%
{(\omega+\Omega)^{4}}\operatorname{Re}\left(  \frac{L}{l}E_{2}^{2}%
(z)-E_{3}(2z)\right)  , \label{I2dirtyexact}%
\end{equation}
where $I_{0}^{\left(  3\right)  }=e\nu_{0}^{\left(  3\right)  }L_{y}L_{z}\pi
v_{F}\Delta_{0}$, $L_{y}L_{z}$ being the junction area, and $z=L/l+i\alpha
=L/l+2iL/\xi_{F}$. The functions $E_{i}(z)$ are defined in the appendix. The
first harmonic corresponds to the result previously obtained in
\cite{buzdin82bis,bergeret01}. The second harmonic is usually smaller than the
first one, except when $I_{1}$ cancels. Then we obtain that the magnitude of
$I_{2}$ is perfectly measurable though small. Moreover, the sign of $I_{2}$
when $I_{1}=0$ is very instructive since it determines the order of the
transition. When the first harmonic cancels, the second harmonic is finite and
positive. This is the usual case where one observes a discontinuous jump of
the junction between the zero and $\pi$ states. For negative $I_{2}$ when
$I_{1}=0$, one would expect a continuous transition and the realisation of a
$\varphi$ junction \cite{buzdin05}. For the experimentally relevant regime of
moderate $L/l$, we always obtain that the second harmonic is positive at the
transition (see Fig. 6). For larger values of $L/l$, we have also observed
negative $I_{2}$ although numerical artefact cannot be excluded.

We now study how the supercurrent is suppressed when the weak link length or
the exchange field is increased. The asymptotic behaviors of the harmonics at
$\alpha\gg1$ are given by%

\begin{equation}
\frac{I_{1}}{I_{0}^{\left(  3\right)  }}=\text{ }T\sum_{\omega>0}\frac
{8\Delta_{0}e^{-L/l}}{(\omega+\Omega)^{2}}\operatorname{Re}\left(
\frac{e^{-2iL/\xi_{F}}}{L/l+2iL/\xi_{F}}\right)  , \label{I1dirty}%
\end{equation}
and
\begin{equation}
\frac{I_{2}}{I_{0}^{\left(  3\right)  }}=T\sum_{\omega>0}\frac{8\Delta_{0}%
^{3}e^{-2L/l}}{(\omega+\Omega)^{4}}\operatorname{Re}\left(  \frac
{(L/l-2iL/\xi_{F})e^{-4iL/\xi_{F}}}{(L/l+2iL/\xi_{F})^{2}}\right)  ,
\label{I2dirty}%
\end{equation}
Besides the exponential suppression $I_{1}\sim e^{-L/l}$ (and $e^{-2L/l}$ for
$I_{2}$), the real parts in Eq.(\ref{I1dirty},\ref{I2dirty}) provide damped
oscillations as a function of $\alpha$, as shown in Fig. 7.

The sum over Matsubara frequencies yields the temperature dependence of the
harmonics. At low temperature $T\rightarrow0$, the sum over Matsubara
frequencies becomes an integral that can be done analytically yielding:
\begin{equation}
\frac{I_{1}}{I_{0}^{\left(  3\right)  }}=\frac{8}{3}\operatorname{Re}\left(
\frac{e^{-2iL/\xi_{F}}}{L/l+2iL/\xi_{F}}\right)  e^{-L/l},
\end{equation}
and
\begin{equation}
\frac{I_{2}}{I_{0}^{\left(  3\right)  }}=\frac{8}{15}\operatorname{Re}\left(
\frac{(L/l-2iL/\xi_{F})e^{-4iL/\xi_{F}}}{(L/l+2iL/\xi_{F})^{2}}\right)
e^{-2L/l},
\end{equation}
at large $\alpha$.

\subsection{\bigskip Two dimensional case.}

\begin{figure}[ptb]
\begin{center}
\includegraphics[width=3.5in]{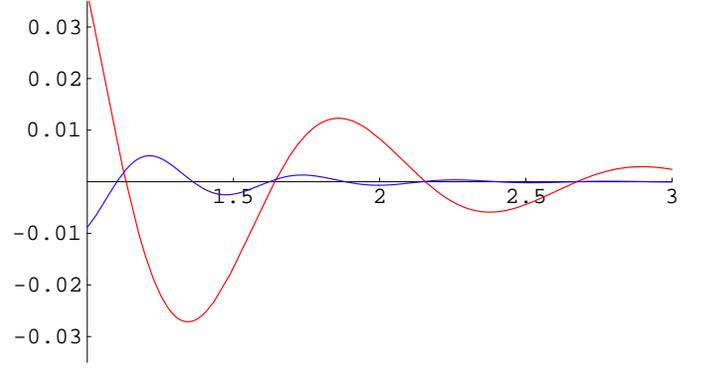}
\end{center}
\caption{First (red curve) and second (blue curve) harmonic as functions of
$L/l$ for a three-dimensional junction. We have chosen $l/\xi_{F}=5$. The
corresponding analytical expressions Eqs.(\ref{I1dirtyexact}%
,\ref{I2dirtyexact}) are derived in the regime $\xi_{F}\lesssim l\lesssim L$.
Hence the lower limit for $L/l$ is one. }%
\label{HarmonicsDirty3D}%
\end{figure}

For planar junctions, both the first and second harmonics were unknown in the
strong ferromagnet regime $\xi_{F}\ll l\ll L$. Using the procedure described
in appendix, we have evaluted these harmonics as
\begin{equation}
I_{1}=16\text{ }I_{0}^{\left(  2\right)  }T\sum_{\omega>0}\frac{\Delta_{0}%
}{(\omega+\Omega)^{2}}\operatorname{Re}F_{2}(z), \label{I1dirtyexact2D}%
\end{equation}
and
\begin{equation}
I_{2}=16\text{ }I_{0}^{\left(  2\right)  }T\sum_{\omega>0}\frac{\Delta_{0}%
^{3}}{(\omega+\Omega)^{4}}\operatorname{Re}\left(  \frac{L}{l}F_{1}%
^{2}(z)-F_{2}\left(  2z\right)  \right)  . \label{I2dirtyexact2D}%
\end{equation}
Here $I_{0}^{\left(  2\right)  }=e\nu_{0}^{\left(  2\right)  }\pi v_{F}%
\Delta_{0}L_{y}$, $L_{y}$ being the width of the planar weak link, and
$z=L/l+2iL/\xi_{F}$. The functions $F_{i}(z)$ are defined in the appendix.
From Fig.\ref{FIG_I_2_2bis_with_tau}, one obseves that the second harmonic is
finite and positive when the first harmonic cancels.

\begin{figure}[ptb]
\begin{center}
\includegraphics[width=3.5in]{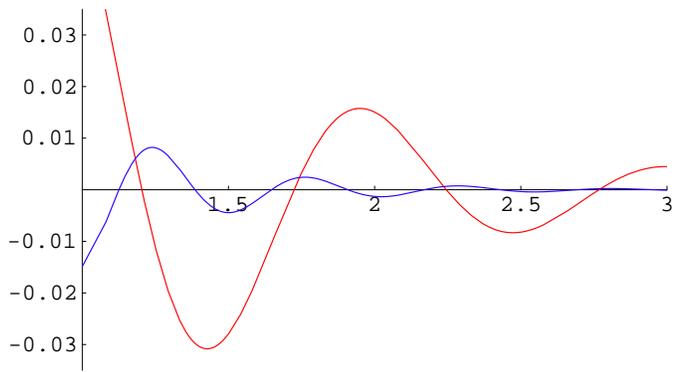}
\end{center}
\caption{First (red curve) and second (blue curve) harmonic as functions of
$L/l$ for a two-dimensional junction. We have chosen $l/\xi_{F}=5$. The
corresponding analytical expressions Eqs.(\ref{I1dirtyexact2D}%
,\ref{I2dirtyexact2D}) are derived in the regime $\xi_{F}\lesssim l\lesssim
L$. Hence the lower limit for $L/l$ is one. }%
\label{FIG_I_2_2bis_with_tau}%
\end{figure}

\subsection{Comparison with experiments}

In the experiments \cite{robinson07}, the values of the parameter $h\tau$ are
respectively $h\tau=3$ for Ni$_{80}$Fe$_{20}$, $2.8$ for Co, $1.62$ for Fe and
$0.5$ for Ni, using the parameters ($h,v_{F},l$) provided in \cite{robinson07}%
. The experiments \cite{blum02} performed on Nb-Ni-Nb junctions correspond to
$h\tau\simeq1$. Hence these experiments cover the onset of the regime
$h\tau\geq1$, or equivalently $\xi_{F}\lesssim l$. The oscillations of the
critical current are reported for weak link lengths $L$ not exceeding few mean
free paths $l$, othewise the signal would be too small (due to the exponential
suppression by the factor $e^{-L/l}$). The period of the oscillations is
approximatively $0.5$ in units of $L/l$ in agreement with our results for
$h\tau=3$, see Fig. 7. Quantitative comparison between our theory and these
experiments are hindered by the fact that band structure effects and interface
quality may strongly influence the magnitude of the Josephson current.

\section{Conclusion.}

In the absence of impurities, we have demonstated that temperature-induced
$0$-$\pi$ transitions are possible both in two and three dimensional SFS
junctions though it requires weak exchange fields in practice. The overall
decay or damping of the critical current as a function of the length/exchange
field of the ferromagnet is much slower ($L^{-1/2}$) in the two-dimensional
case than in the three dimensional case ($L^{-1}$). Moreover, the shape of the
critical current versus temperature curves changes when one closely approaches
a $0$-$\pi$ transition. Hence for future experiments, we suggest the
challenging measure of the temperature dependence of a minimum of the critical
current, which is essentially given by the contributions of higher harmonics
($m\geqslant2$). The corresponding $I_{c}(T)$ should differ markedly form the
usual almost linear curves obtained so far away from those minima
\cite{robinson07,blum02}.

We have obtained the current phase relation for SFS junctions comprising
strong ferromagnets in the presence of moderate disorder, namely in the limit
$\tau h\gtrsim1$. We have calculated the second harmonic, in particular at the
$0$-$\pi$ transition, for both two- and three- dimensional SFS junctions. In
the three dimensional case, we have compared our result with recent
experiments performed in the regime $\tau h\gtrsim1$ \cite{robinson07}.

\bigskip

\section{Appendix.\bigskip}

In this appendix, we derive the current-phase relation in the regime $L\gg
\ell\gg\xi_{F},$ or equivalently $\alpha\gg\tau h\gg1$. We start with the
normal quasiclassical Green function which is uniform in the ferromagnet:%

\begin{equation}
g_{\omega}\left(  \theta\right)  =\frac{\omega}{\Omega}+\frac{\Delta_{0}^{2}%
}{\Omega}\frac{\sinh\Phi(\theta)}{\omega\sinh\Phi(\theta)\pm\Omega\cosh
\Phi(\theta)},
\end{equation}
where $+/-$ corresponds to the sign of $\cos\theta$ and
\begin{equation}
2\Phi(\theta)=i\chi+\frac{\left\langle g\right\rangle L/l+i\alpha}{\cos\theta}%
\end{equation}
with $\alpha=2hL/v_{F}$. We first consider the case of positive Matsubara
frequency $\omega>0$. Then, one expands $g_{\omega}\left(  \theta\right)  $ in
powers of $X_{\lambda}=\exp\left[  -2\lambda\Phi(\theta)\right]  $ as
\begin{equation}
g_{\omega}\left(  \theta\right)  =1-\frac{2\Delta_{0}^{2}}{(\omega+\Omega
)^{2}}X_{\lambda}+\frac{2\Delta_{0}^{4}}{(\omega+\Omega)^{4}}X_{\lambda}%
^{2}+...
\end{equation}
where $\lambda$ is the sign of $\cos\theta$. The modulus of $X_{\lambda}$ is a
small parameter because $\left\vert X_{\lambda}\right\vert =\exp
(-2\lambda\left\langle g\right\rangle L/(l\cos\theta))$ with $\left\langle
g\right\rangle >0$ (In first approximation $\left\langle g\right\rangle
=sgn(\omega)$) and $L\gg l$. This expansion defines a self-consistent problem
since $X_{\lambda}$ contains $\left\langle g\right\rangle $ which in turn is
evaluated using $X_{\lambda}$. As a first iteration, we evaluate $g_{\omega
}\left(  \theta\right)  $ using $\left\langle g\right\rangle =1$ in the
expression of $X_{\lambda}$. Then we perform the angular averaging to obtain
the first order correction to $\left\langle g\right\rangle $. Denoting
$\left\langle g\right\rangle =1+g_{1}$ ( $\left\vert g_{1}\right\vert \ll
1$)\textit{ }one obtains, \textit{in the three dimensional case: }%

\begin{align}
g_{1}  &  =-\frac{2\Delta_{0}^{2}\cos\chi}{(\omega+\Omega)^{2}}%
{\displaystyle\int\limits_{0}^{\pi/2}}
d\theta\sin\theta\exp\left(  -\frac{z}{\cos\theta}\right) \\
&  =-\frac{2\Delta_{0}^{2}\cos\chi}{(\omega+\Omega)^{2}}E_{2}(z)
\end{align}
with $z=L/l+i\alpha$ and $E_{n}(z)=%
{\displaystyle\int\limits_{1}^{\infty}}
dy$ $y^{-n}e^{-zy}$.

The current is related to the quantity:
\begin{align}
\left\langle v_{x}g_{\omega}\left(  \theta\right)  \right\rangle  &
=-\frac{v_{F}\Delta_{0}^{2}}{(\omega+\Omega)^{2}}%
{\displaystyle\int\limits_{0}^{\pi/2}}
d\theta\sin\theta\cos\theta(X_{\oplus}-X_{\ominus})\label{VG}\\
&  +\frac{v_{F}\Delta_{0}^{4}}{(\omega+\Omega)^{4}}%
{\displaystyle\int\limits_{0}^{\pi/2}}
d\theta\sin\theta\cos\theta(X_{\oplus}^{2}-X_{\ominus}^{2})\nonumber
\end{align}
In this expression, the first integral (proportional to $\Delta_{0}^{2}$) must
be evaluated within the approximation:
\begin{equation}
X_{\oplus}-X_{\ominus}=-2i\sin\chi\exp\left(  -\frac{z}{\cos\theta}\right)
\left(  1-g_{1}\frac{L/l}{\cos\theta}\right)  , \label{X1}%
\end{equation}
whereas the zero-order approximation $\left\langle g\right\rangle =1:$%
\begin{equation}
X_{\oplus}^{2}-X_{\ominus}^{2}=-2i\sin2\chi\exp\left(  -\frac{2z}{\cos\theta
}\right)  \label{X2}%
\end{equation}
is sufficient for the second integral (proportional to $\Delta_{0}^{4}$).
Substituing Eq.(\ref{X1},\ref{X2}) in Eq.(\ref{VG}) yields:
\begin{align*}
\left\langle v_{x}g_{\omega}\left(  \theta\right)  \right\rangle  &
=\frac{2iv_{F}\Delta_{0}^{2}}{(\omega+\Omega)^{2}}E_{3}(z)\sin\chi\\
&  +\frac{2iv_{F}\Delta_{0}^{4}}{(\omega+\Omega)^{4}}\left(  \frac{L}{l}%
E_{2}^{2}(z)-E_{3}(2z)\right)  \sin2\chi\text{ }%
\end{align*}
and finally Eq.(\ref{I1dirtyexact},\ref{I2dirtyexact}).

\textit{In the two dimensional case,} the first order correction to
$\left\langle g\right\rangle $ reads: \textbf{ }
\begin{align}
g_{1}  &  =-\frac{2\Delta_{0}^{2}\cos\chi}{\pi(\omega+\Omega)^{2}}%
{\displaystyle\int\limits_{0}^{\pi/2}}
d\theta\exp\left(  -\frac{L/l+i\alpha}{\cos\theta}\right) \\
&  =-\frac{2\Delta_{0}^{2}\cos\chi}{(\omega+\Omega)^{2}}F_{1}(z)\text{,}%
\end{align}
where:
\begin{equation}
F_{n}(z)=%
{\displaystyle\int\limits_{1}^{\infty}}
\frac{dye^{-zy}}{\pi y^{n}\sqrt{y^{2}-1}}.
\end{equation}
The current is related to the average:
\begin{align*}
\left\langle v_{x}g_{\omega}\left(  \theta\right)  \right\rangle  &  =v_{F}%
{\displaystyle\int\limits_{0}^{\pi/2}}
\frac{d\theta}{\pi}\cos\theta\left[  g(\theta)-g(\pi-\theta)\right] \\
&  =-\frac{2v_{F}\Delta_{0}^{2}}{(\omega+\Omega)^{2}}%
{\displaystyle\int\limits_{0}^{\pi/2}}
\frac{d\theta}{\pi}\cos\theta(X_{\oplus}-X_{\ominus})\\
&  +\frac{2v_{F}\Delta_{0}^{4}}{(\omega+\Omega)^{4}}%
{\displaystyle\int\limits_{0}^{\pi/2}}
\frac{d\theta}{\pi}\cos\theta(X_{\oplus}^{2}-X_{\ominus}^{2}).
\end{align*}
Proceeding to the same approximations as for the three-dimensional case, one obtains:%

\begin{align*}
\left\langle v_{x}g_{\omega}\left(  \theta\right)  \right\rangle  &
=\frac{4i\sin\chi v_{F}\Delta_{0}^{2}}{(\omega+\Omega)^{2}}F_{2}%
(z)\mathbf{+}\\
&  +\frac{4i\sin2\chi v_{F}\Delta_{0}^{4}}{(\omega+\Omega)^{4}}\left(
\frac{L}{l}F_{1}^{2}(z)-F_{2}\left(  2z\right)  \right)  \text{ }%
\end{align*}
and finally Eqs.(\ref{I1dirtyexact2D},\ref{I2dirtyexact2D}).

\bigskip

\begin{acknowledgments}
The authors acknowledge Takis Kontos, Julien Morthomas and Joseph Leandri for
helpful discussions. This work was supported by the Agence Nationale de la
Recherche Grant No. ANR-07-NANO-011: ELEC-EPR.
\end{acknowledgments}


\bigskip

\begin{references}
\bibitem{josephson62} B. Josephson, Phys. Lett. 1, 251–253 ( 1962).
\bibitem{golubov04} A. Golubov, M. Kupriyanov, and E. Il'ichev, Rev. Mod. Phys. 76, 411 (2004).
\bibitem{bulaevskii78} L. N. Bulaevskii, V. V. Kuzii, and A. A. Sobyanin, Pis'ma Zh.Éksp. Teor. Fiz. 25, 314 (1977) [JETP Lett. 25, 290 (1977)].
\bibitem{bulaevskii78ss}  L.N. Bulaevskii, V.V. Kuzii, and A.A. Sobyanin, Solid State Commun. {\bf 25}, 1053 (1978).
\bibitem{kulik66} I.O. Kulik,  JETP {\bf 22}, 841 (1966).
\bibitem{shiba69} H. Shiba and T. Soda, Prog. Theor. Phys., {\bf 41}, 25, (1969).
\bibitem{buzdin82}   A.I. Buzdin, L.N. Bulaevskii, and S.V. Panyukov, JETP Lett. {\bf 35}, 178 (1982).
\bibitem{buzdin82bis}   A.I. Buzdin, L.N. Bulaevskii, and S.V. Panyukov, Solid State Commun. {\bf 44}, 539 (1982).
\bibitem{ryazanov01} V. V. Ryazanov, V. A. Oboznov, A. Yu. Rusanov, A. V. Veretennikov, A. A. Golubov, and J. Aarts, Phys. Rev. Lett. 86, 2427 (2001).
\bibitem{kontos02} T. Kontos, M. Aprili, J. Lesueur, F. Genet, B. Stephanidis, and R. Boursier, Phys. Rev. Lett. {\bf 89}, 137007 (2002).
\bibitem{buzdin91}   A.I. Buzdin and M.Y. Kuprianov, JETP Lett. {\bf 53}, 308 (1991).
\bibitem{kopnin} N.B. Kopnin, in Theory of Nonequilibrium Superconductivity. The International Series of Monographs on Physics, Vol. 110 (Clarendon Oxford, 2001).
\bibitem{likharev79} K.K. Likharev, Rev. Mod. Phys. {\bf 51}, 101 (1979).
\bibitem{buzdin05} A. I. Buzdin, Rev. Mod. Phys. 77, 935 (2005).
\bibitem{lyuksyutov05} I. F. Lyuksyutov and V. L. Pokrovsky, Adv. Phys. 54, 67 (2005).
\bibitem{guichard03} W. Guichard, M. Aprili, O. Bourgeois, T. Kontos, J. Lesueur, and P. Gandit, Phys. Rev. Lett. {\bf  90}, 167001 (2003).
\bibitem{bauer04} A. Bauer, J. Bentner, M. Aprili, M.L. Della Rocca, M. Reinwald, W. Wegscheider, and C. Strunk, Phys. Rev. Lett. {\bf 92}, 217001 (2004).
\bibitem{ryazanovNatPhys} S.M. Frolov, M.J.A. Stoutimore, T.A. Crane, D.J. Van Harlingen, V.A. Oboznov, V.V. Ryazanov, A. Ruosi, C. Granata, and M. Russo, Nature Physics {\bf 4}, 32 (2008).
\bibitem{oboznov06} V. A. Oboznov, V. V. Bolginov, A. K. Feofanov, V. V. Ryazanov, and A. I. Buzdin, Phys. Rev. Lett {\bf 96}, 197003 (2006).
\bibitem{robinson07} J. W.A. Robinson, S. Piano, G. Burnell, C. Bell, and M. G. Blamire, Phys. Rev. B {\bf 76}, 094522 (2007).
\bibitem{frolov04} S.M. Frolov, D.J. Van Harlingen, V.A. Oboznov, V.V. Bolginov, and V.V. Ryazanov, Phys. Rev. B {\bf 70}, 144505 (2004).
\bibitem{dellarocca07} M.L. Della-Rocca, M. Chauvin, B. Huard, H. Pothier, D. Esteve and C. Urbina, Phys. Rev. Lett. {\bf 99}, 127005 (2007).
\bibitem{troeman08} A. G.P. Troeman, S. H. van der Ploeg, E. Il'Ichev, H.-G. Meyer, A. A. Golubov, M. Yu. Kupriyanov, and H. Hilgenkamp, Phys. Rev. B {\bf 77}, 024509 (2008).
\bibitem{strunk08} M. Fuechsle, J. Bentner, D.A. Ryndyk, M. Reinwald, W. Wegscheider, and C. Strunk, arXiv0707.4512
\bibitem{sellier04}  Hermann Sellier, Claire Baraduc, François Lefloch, and Roberto Calemczuk , Phys. Rev. Lett. {\bf 92}, 257005 (2004).
\bibitem{buzdin05prb} A. Buzdin, Phys. Rev. B {\bf 72}, 100501(R) (2005).
\bibitem{houzet05} M. Houzet, V. Vinokur, and F. Pistolesi, Phys. Rev. B {\bf 72}, 220506(R) (2005).
\bibitem{zareyan06} G. Mohammadkhani and M. Zareyan, Phys. Rev. B {\bf 73}, 134503 (2006).
\bibitem{blum02} Y. Blum, A. Tsukernik, M. Karpovski, and A. Palevski, Phys. Rev. Lett. {\bf 89}, 187004 (2002).
\bibitem{robinson06} J. W.A. Robinson, S. Piano, G. Burnell, C. Bell, and M. G. Blamire, Phys. Rev. Lett. {\bf 97}, 177003 (2006).
\bibitem{bergeret01} F. S. Bergeret, A. F. Volkov, and K. B. Efetov, Phys. Rev. B {\bf 64}, 134506 (2001).
\bibitem{uchoa08} B. Uchoa, C-Y. Lin, and A.H. Castro Neto, Phys. Rev. B {\bf 77}, 035420 (2008).
\bibitem{uchoa07} B. Uchoa and A.H. Castro Neto, Phys. Rev. Lett. {\bf 98}, 146801 (2007).
\bibitem{heersche07} H.B. Heersche, P. Jarillo-Herrero, J.B. Oostinga, L.M.K. Vandersypen and A.F. Morpurgo, Nature {\bf 446} , 56 (2007).
\bibitem{du08} Xu Du, Ivan Skachko, and Eva Y. Andrei , Phys. Rev. B {\bf 77}, 184507 (2008).
\bibitem{huertas08} J. Linder, T. Yokoyama, D. Huertas-Hernando, and A. Sudbo, Phys. Rev. Lett. {\bf 100}, 187004 (2008).
\bibitem{keiser06} R.S. Keiser, S.T.B Goennenwein, T.M. Klapwijk, G. Miao, G. Xiao, and A. Gupta, Nature {\bf 439}, 825 (2006).
\bibitem{bergeret05} F. S. Bergeret, A. F. Volkov, and K. B. Efetov, Rev. Mod. Phys. {\bf 77}, 1321 (2005).
\bibitem{cayssol04} J. Cayssol and G. Montambaux, Phys. Rev. B {\bf 70}, 224520 (2004).
\bibitem{cayssol05} J. Cayssol and G. Montambaux, Phys. Rev. B {\bf 71}, 012507 (2005).
\bibitem{cottet05} A. Cottet and W. Belzig, Phys. Rev. B {\bf 72}, 180503(R) (2005).
\end{references}
\end{document}